This article appeared in Rev. Sci. Instrum. 96(4), 044707 (2025). DOI: 10.1063/5.0252515.

# Modeling the differential susceptibility by Lorentzians

Alexej Perevertov

**AFFILIATION**

FZU - Institute of Physics of the Czech Academy of Sciences, Department of Magnetic Measurements and Materials, 18200 Prague, Czech Republic

## ABSTRACT

The idea to extract information on magnetically different phases from magnetic measurements is very attractive and many efforts have been made in this area. One of the most popular direction is to use the Preisach model formalism to analyze the 2D Preisach distribution function (PDF) obtained either from first order reversal curves (FORC) or minor loops. Here we present an alternative a much simpler procedure – the analysis of the derivative of the saturation magnetization loop, the differential susceptibility curve. It follows the Lorentzian shape with very high accuracy for ferromagnetic polycrystalline materials. This allows decomposing any differential susceptibility curve of a complex multi-phase material into individual components representing different magnetic phases by Lorenzian peaks – in the same way as it is done in X-ray diffraction analysis of materials. We show that the minor differential susceptibility curves also have the Lorenzian shape that can facilitate calculation of the Preisach distribution function from the experimental curves and reduce noise in the resulting PDF.

## I. Introduction

Magnetic hysteresis loop, *M(H)* loop shows the relationship between magnetization, *M* and the applied magnetic field, *H*. It is a subject of intensive study for the last 150 years since it characterizes magnetic properties of a material. The *M(H)* loop shape changes with the field amplitude increase – the coercive force, $H_C$ and the maximum differential susceptibility (DS) increase. After reaching a certain value of the field amplitude, $H_C$ and the shape of DS do not change anymore [1]. This moment is called the technical saturation. With the further field increase the magnetization grows in a fully reversible way, mostly by the magnetic moments rotation from the crystal magnetic easy axes toward the applied field direction. The field necessary to orient all the magnetic moments in a polycrystalline material is at least two orders larger comparing to the technical saturation field. In most cases investigators work with the technical saturation *M(H)* loops, called major hysteresis loops. The loops with smaller amplitudes are called “minor loops” and their shape strongly depends on the magnetic prehistory.

The modeling of *M(H)* hysteresis loops is crucial not only for practical applications but also for fundamental research, as it enables a deeper understanding of magnetic property evolution under variations in composition, temperature, stress, and other factors. Hundreds of studies have been dedicated to hysteresis loop modeling [2–10]. An effective model should be as simple as possible while accurately reproducing experimental magnetization curves. Ideally, it should be expressed as a straightforward analytical function with a minimal number of parameters.

Currently, the most widely recognized and commonly used hysteresis models are the Preisach and Jiles–Atherton models, both of which are relatively complex. Additionally, several new models have been introduced in recent years, some involving a large number of parameters and significant computational complexity [5,11–12]. Over the past decades, various analytical functions have been explored for representing major hysteresis loops, anhysteretic magnetization curves, and two-dimensional Preisach distribution functions (PDFs) [13–25].

A typical *M(H)* curve exhibits an "S" shape, a characteristic of mathematical functions known as sigmoid functions. Numerous analytical sigmoid functions closely resemble experimentally measured hysteresis loops. For a single-phase magnetic material, a key criterion for an appropriate hysteresis model is the correct asymptotic behavior, ensuring saturation at infinitely large fields. The differential susceptibility (DS), which is the first derivative of the hysteresis curve, should exhibit a single peak for one-phase magnetic materials.

Various sigmoid functions can be used for modeling hysteresis, including the arctangent function, the hyperbolic tangent, the Gudermannian function, the error function, the generalized logistic function, and algebraic functions, among others. Examples of several simple sigmoid functions are shown in Fig. 1, all of which approximate real experimental *M(H)* curves. If a deviation of 5–10% from the experimental magnetization curve is acceptable, any of these functions can be effectively used for magnetic hysteresis modeling. However, these functions differ in their approach to saturation; for example, the arctangent and Langevin functions exhibit significantly slower saturation behavior. Notably, the derivative of an idealized rectangular *M(H)* loop corresponds to a delta function.

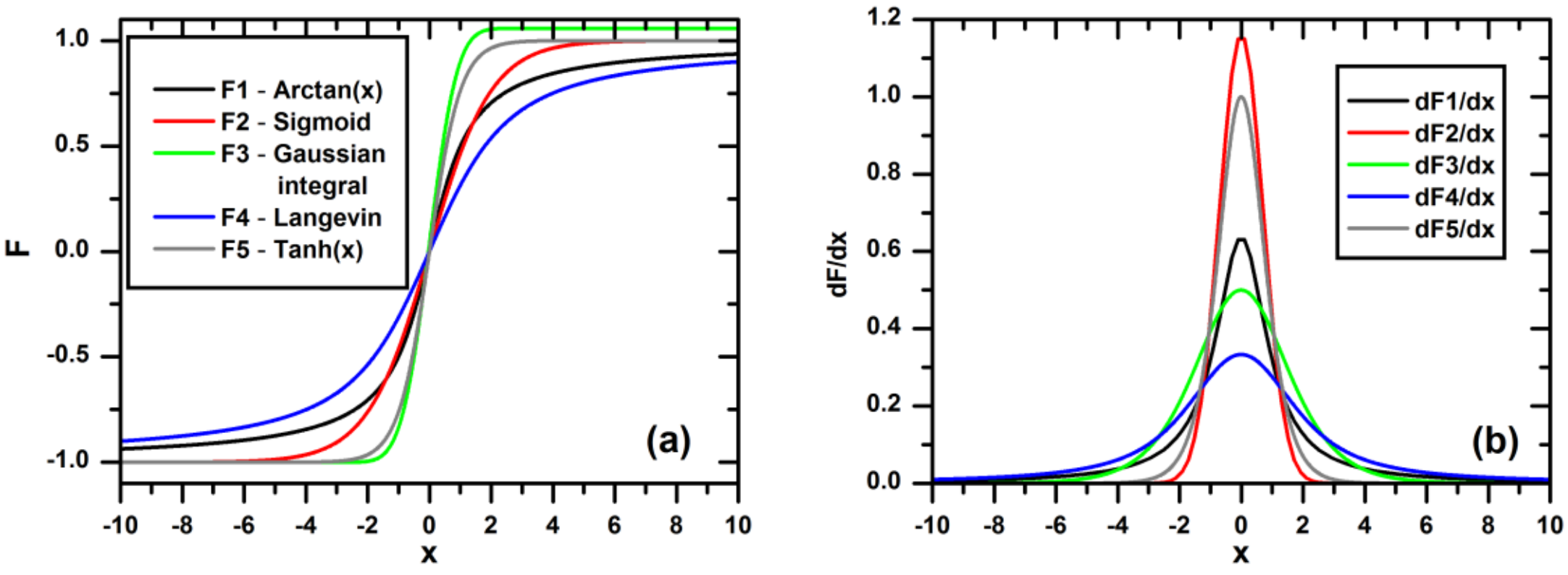

(a) (b)

**FIG.1.** Several analytical functions representing ascending magnetization curves (a) and their derivatives - the differential susceptibility. F1 is the arctangent function, F2 – the sigmoid function, F3 – the integral of Gaussian, F4 – the Langevin function and F5 – the hyperbolic tangent. All the functions are scaled and biased to the ±1 magnetization range.

In the 1990s, many researchers anticipated that the Preisach model formalism would be applied similarly to Fourier analysis to extract information about individual magnetic phases within a material. It was hypothesized that the two-dimensional Preisach distribution function (PDF) would exhibit additional peaks compared to traditional susceptibility curves. However, our findings indicate that this expectation was not realized. Specifically, for a differential susceptibility curve with several peaks, the corresponding PDF also contains the same number of peaks.

The additional information that can be extracted from minor hysteresis loops or first-order reversal curves (FORC) pertains primarily to magnetostatic interactions between magnetic phases, which manifest as negative regions in the PDF [26]. We found that direct analysis of a set of minor susceptibility curves is sufficient to detect even the weakest interactions, eliminating the need for explicit PDF calculations [26]. Experimentally obtained PDFs are often highly noisy, and their measurement and evaluation require complex procedures. Moreover, we have demonstrated that the PDF can be derived directly from differential susceptibility (DS) curves without first computing the integral quantity, i.e., the *M(H)* loops [26]. Consequently, we have focused our investigations on DS curves rather than PDFs, as DS curves already contain essential information regarding the number of magnetic phases in a material. Within the Preisach model, the DS curve of a saturated hysteresis loop corresponds to the distribution of switching-up fields.

Most studies either analyze *M(H)* curves or examine the second derivative, i.e., the PDF function. However, relatively few publications have explored the characteristics of the first derivative, the DS curve. This article aims to address this gap. Even when differences between *M(H)* curves are imperceptible to the naked eye, the corresponding differential susceptibility curves may exhibit significant variations. The DS curve, being the first derivative, is substantially more sensitive than the *M(H)* curve while also being significantly less noisy than the second derivative, the PDF. Thus, the shape of the DS curve can be studied in detail, providing valuable insights into the magnetic properties of materials.

## II. Lorentzian shape of the differential susceptibility of polycrystalline materials

We have measured a large number of experimental *M(H)* curves of polycrystalline materials like steels for the last twenty years. In most cases, we worked directly with the DS curves. We have noticed from the beginning that the Lorentzian function always described the DS curves of steels with very high accuracy, accordingly the *M(H)* curves had an arctangent shape [27]. It perfectly worked in all cases – for single-phase materials with one DS peak, for materials with a second magnetically different layer, and even for samples with two-peak DS due to stress-induced anisotropy. It worked for steels with the grain size from tens of microns to several centimeters, for different materials like pure iron, pure nickel, cast irons (see Fig.2 and Table 1), and magnetic shape memory alloys. These observations suggest that it is a general law for the magnetization curve of polycrystalline materials to have the Lorentzian shape of the DS curve. As a result, the DS curves of polycrystalline ferromagnetic materials could be described with very high accuracy by a sum of Lorentzians and so the magnetic spectroscopy can be introduced based on the decomposition of the DS curve of a material into Lorentzian peaks in the same way as it is commonly done in X-ray diffraction analysis of materials.

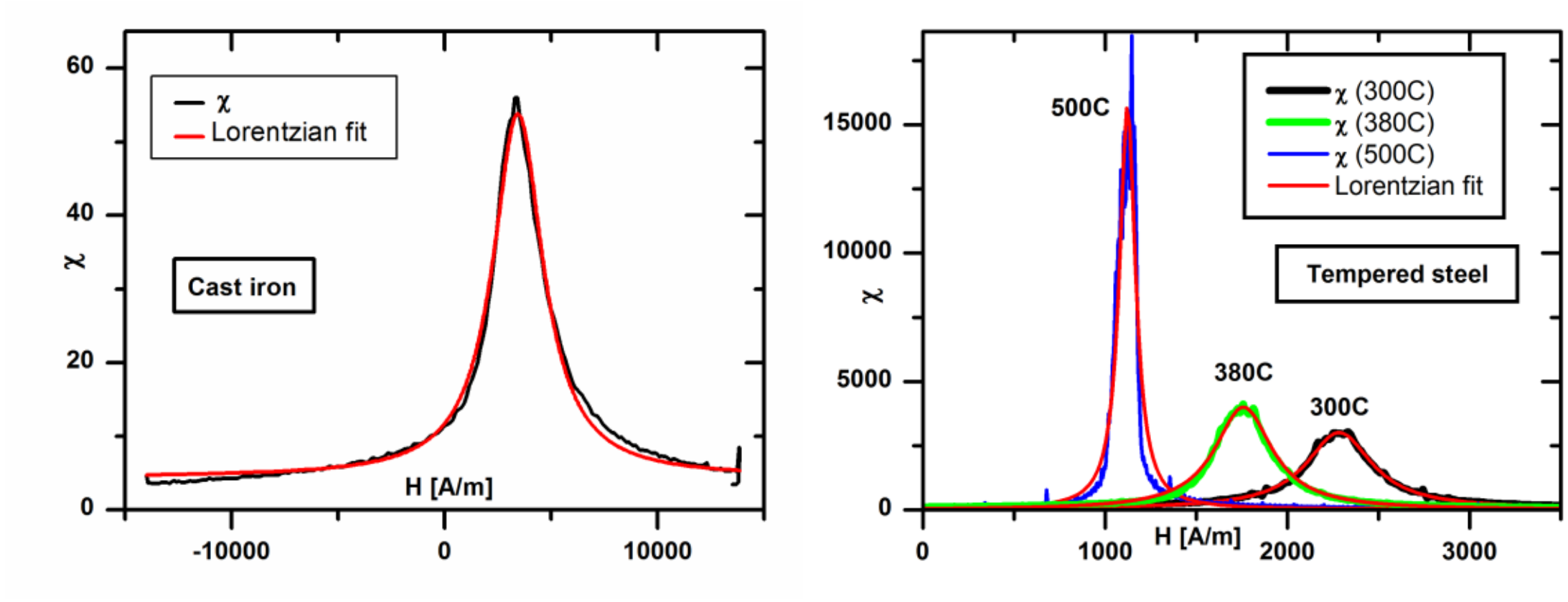

(a) (b)

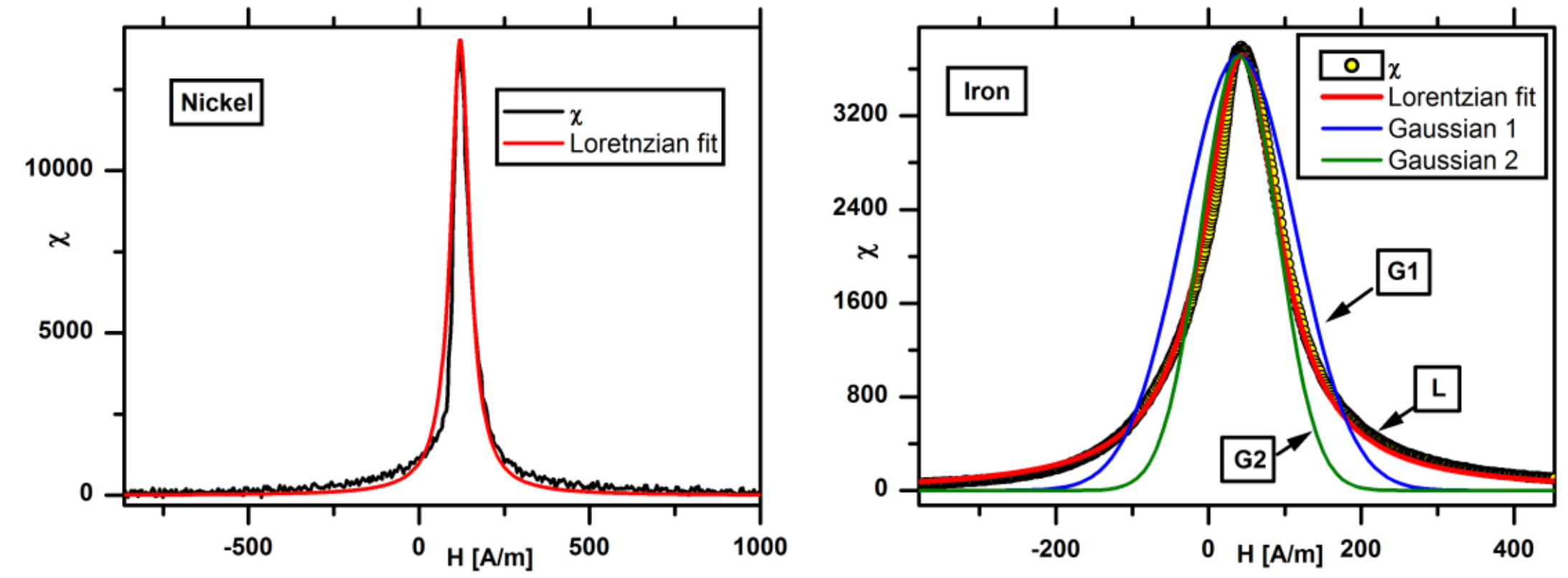

(c) (d)

**FIG.2.** Differential susceptibility curves, $\chi(H)$ and corresponding Lorentzian fits of cast iron (a), quenched and tempered for one hour at 300°C, 380°C and 500°C spring steels (b) [s1], pure nickel (c) and pure iron (d). In (d) also two Gaussian fits are shown. The Gaussian 1 – the same area under the curve and the maximum as of the experimental susceptibility curve. The Gaussian 2 – the same maximum value and the same half-width.

The arctangent function for the ascending branch of the $M(H)$ loop and its derivative – the Lorentzian are given below:

$$M(H) = (2/\pi)\, M_{SAT} \arctan(a[H-H_C])$$

$$\chi(H) = dM/dH = (2/\pi)\, a\, M_{SAT} / (1 + a^2(H-H_C)^2), \quad (1)$$

Where $M_{SAT}$ is the saturation magnetization, $H$ – internal magnetic field, $\chi$ - the differential susceptibility (DS), $H_C$ – the coercive field.

The Lorentzian parameters for materials presented in Fig.2 are given in the Table 1.

Table 1. Lorentzian parameters for different materials (see Fig.2).

| *Material* | $(2/\pi)\, a\, M_{MAX}$ | $1/a$ (A/m) | $H_C$ (A/m) |
|---|---|---|---|
| Cast iron | 53 | 1290 | 3450 |
| Tempered steel, 300°C | 3025 | 223 | 2282 |
| Tempered steel, 380°C | 4000 | 192 | 1760 |
| Tempered steel, 500°C | 15500 | 59 | 1127 |
| Nickel | 14060 | 33 | 120 |
| Iron | 3720 | 61 | 43 |

In Fig.3 the DS curves for Goss-textured Fe-3%Si steel sheets without stress and under applied elastic compressive stress of 45MPa and 67MPa are shown. The DS of compressed samples was modeled by a sum of two Lorentzians. It is seen that the agreement with experimental curves is very good. The average grain size was about 10 mm, the details of measurements can be found in [28]. The agreement was very good for all stresses above 5-10 MPa. Below 5 MPa there is a transition from one-peak to two-peak DS and so the DS curve is the combination of the zero-stress one-peak and two peaks by the domain reorientations. Two peaks in DS are due to stress-induced anisotropy by compression, each of them also having the Lorentzian shape.

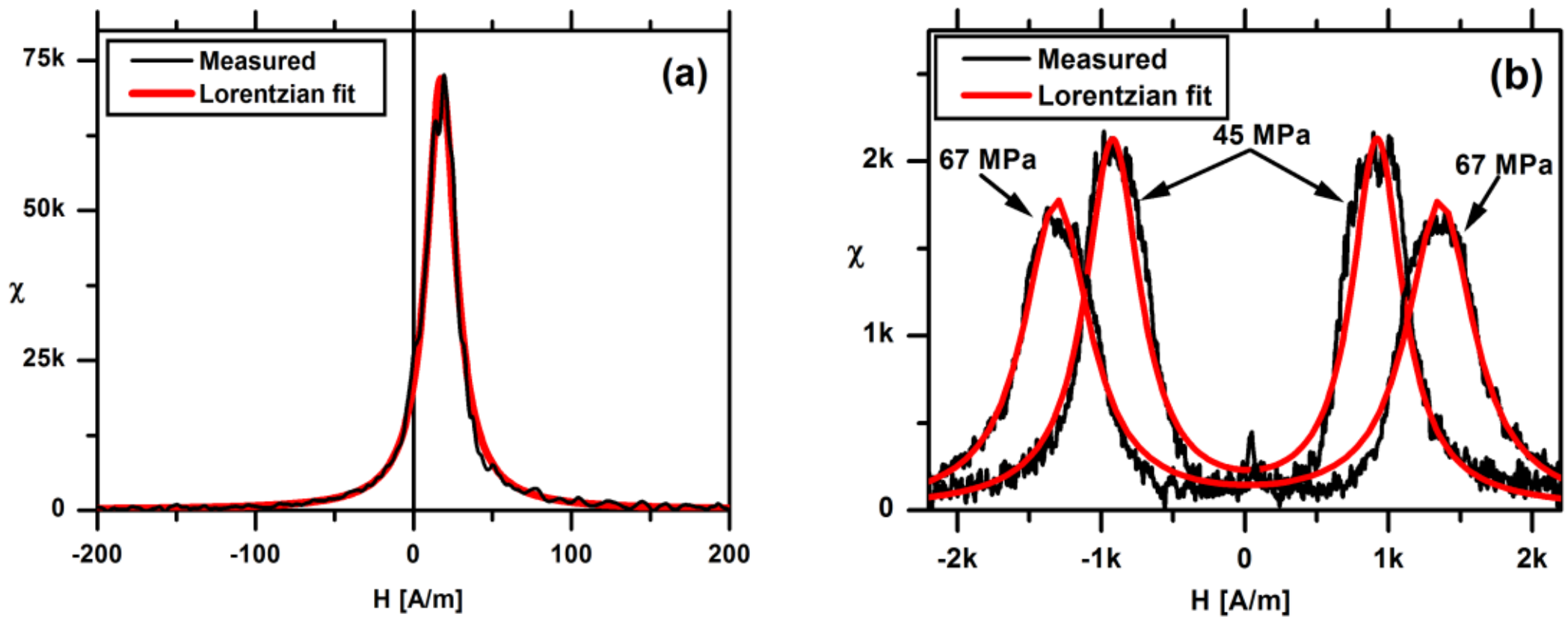

**FIG.3.** Differential susceptibility curves $\chi$ *(H)* for the Goss textured Fe-3%Si steel without stress (a) and under 45MPa and 67MPa compressive stress (b).

Table 2. Lorentzian parameters for grain oriented steel under compression (see Fig.3).

| $\sigma$, *MPa* | $(2/\pi)\, a_1 M_{MAX}$ | $(2/\pi)\, a_2 M_{MAX}$ | *1/a1* *(A/m)* | *1/a2* *(A/m)* | $H_{C1}$, *A/m* | $H_{C2}$, *A/m* |
|---|---|---|---|---|---|---|
| 0 | 72000 | -- | 11.2 | -- | 17.5 | -- |
| 45 | 2100 | 2100 | 182 | 167 | 904 | 924 |
| 67 | 3000 | 3000 | 222 | 200 | 1332 | 1380 |

The differential susceptibility (DS) of plastically deformed low-carbon steel, as reported in our previous work [29], can also be accurately modeled as a sum of two Lorentzian functions (see Fig. 4 and Table 3). Additionally, in that study, several samples were machined to achieve a flatter surface, thereby minimizing air gaps between the sample and the magnetizing yoke. However, this process resulted in the formation of an unintended thin decarburized surface layer, which exhibited significantly softer magnetic properties compared to the bulk material.

The DS curve of these modified samples consists of three distinct peaks: two corresponding to bulk material, attributed to magnetic domain reorientations induced by stress anisotropy [28] (see Fig. 3(b)), and a third peak originating from the thin decarburized surface layer. Figure 5 presents the experimental DS curve of such a sample, modeled as a sum of three Lorentzian functions. The agreement between the experimental data and the model is very good, with the measured and fitted DS curves being virtually indistinguishable. The Lorentzian parameters for three peaks are given in the Table 4.

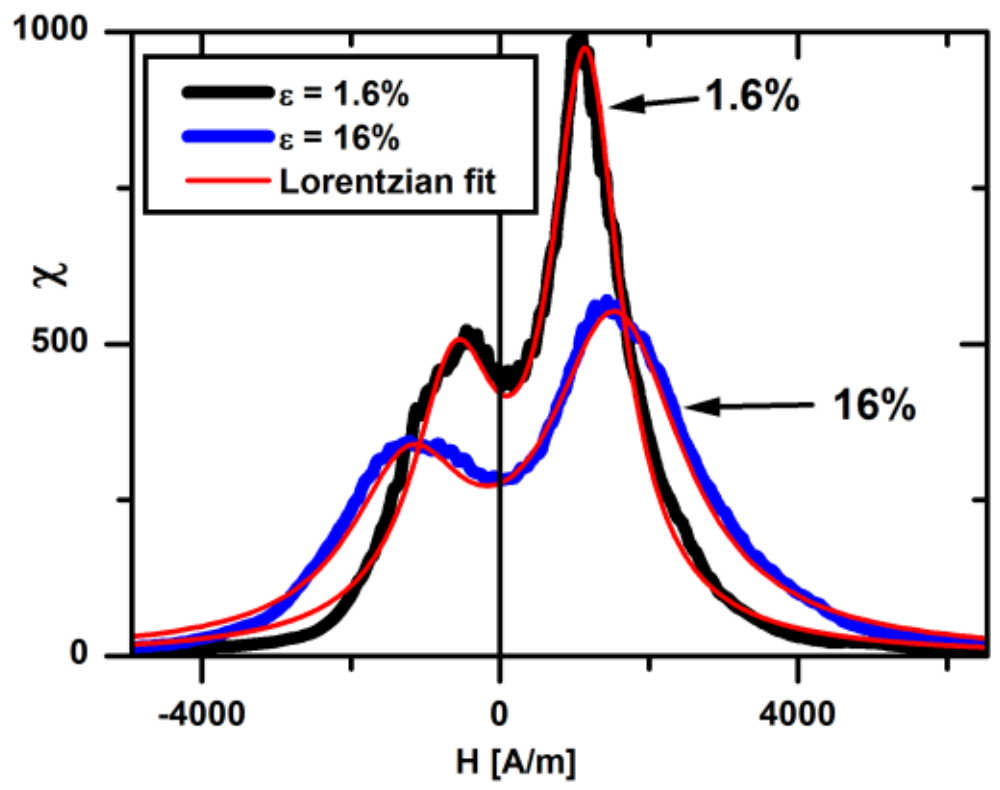

**FIG.4.** Differential susceptibility curves $\chi$ *(H)* of plastically deformed low-carbon steel and two-Lorenzians fit.

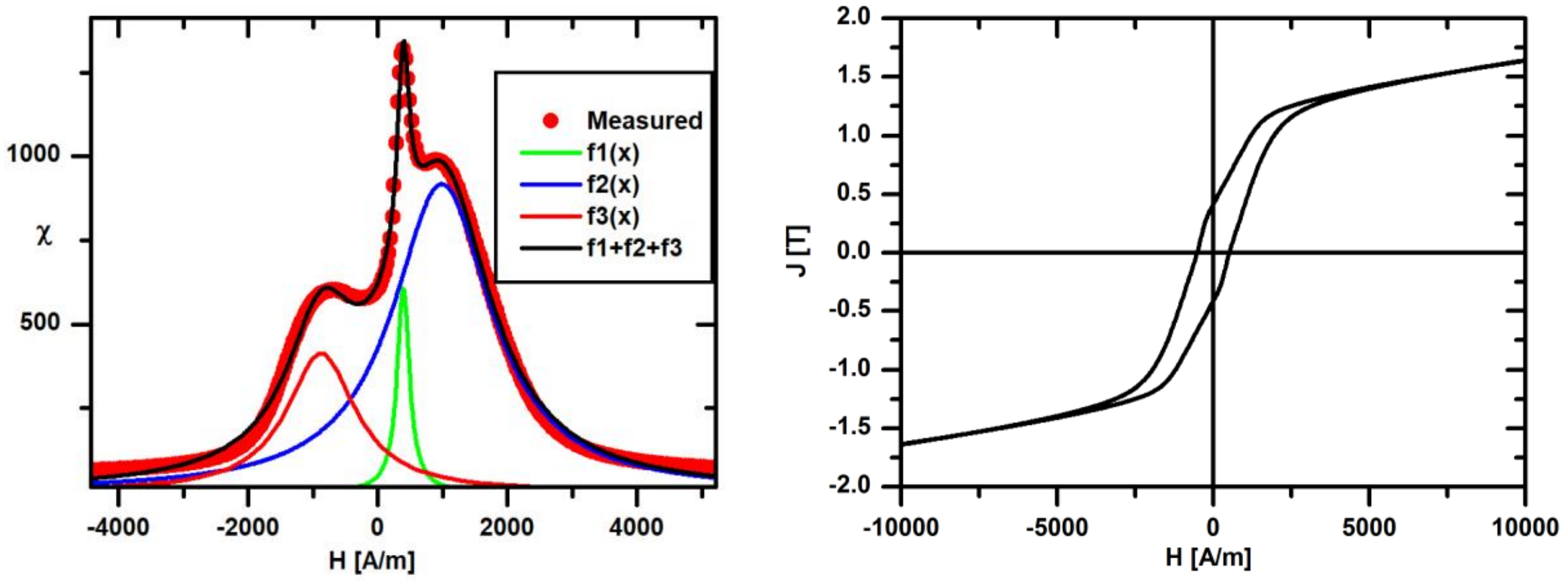

(a) (b)

**FIG.5.** Three-peaks $\chi(H)$ (a) for a deformed low-carbon steel [29] and corresponding magnetic hysteresis loop (b) measured along the hard axis. Two peaks at 1000 and -900 A/m are due to magnetic domains reorientation from the field direction to that perpendicular to it and back. The third thin peak is given by a thin magnetically soft decarburized layer. The measured curve is represented as a sum of three Lorentzians: $F(x) = f1(x)+f2(x)+f3(x)$.

Table 3. Lorentzian parameters for two DS peaks for plastically deformed steel (see Fig.4).

| $\varepsilon$, % | $(2/\pi)\, a_1 M_{MAX}$ | $(2/\pi)\, a_2 M_{MAX}$ | *1/a1* *(A/m)* | *1/a2* *(A/m)* | $H_{C1}$, *A/m* | $H_{C2}$, *A/m* |
|---|---|---|---|---|---|---|
| 1.6 | 916 | 417 | 570 | 700 | 1150 | -600 |
| 16 | 525 | 267 | 1550 | -1220 | 1550 | -1220 |

Table 4. Lorentzian parameters for three DS peaks in Fig.5.

| *Peak* | $(2/\pi)\, a\, M_{MAX}$ | *1/a* (A/m) | $H_C$ (A/m) |
|---|---|---|---|
| 1 | 612 | 83 | 394 |
| 2 | 917 | 665 | 988 |
| 3 | 412 | 467 | -873 |

Let us discuss now the applicability of the model to minor curves. A natural question arises: do minor DS curves also exhibit a Lorentzian shape? Figure 6 presents a set of minor DS curves measured on low-carbon steel (ČSN 12014) after demagnetization. These minor DS curves retain a Lorentzian shape with a consistent width (see Table 5). However, they exhibit asymmetry, as the applied magnetic field range is symmetric with respect to zero field rather than the minor loop coercivity ($H_C$). According to the Preisach model, the terminal points of the minor curves should reconstruct the virgin magnetization curve [30].

Fitting minor DS curves with Lorentzian functions can significantly reduce noise in the calculated Preisach distribution function (PDF) used in Preisach modeling. The observed shift of minor DS curves with increasing field amplitude is a common characteristic in steels and is associated with magnetostatic interactions between different material regions, a phenomenon typically treated within the Preisach framework as a mean-field effect [26]. The dependence of DS curve position and coercivity ($H_C$) on the maximum applied field ($H_{MAX}$) provides valuable insights into magnetostatic interactions between distinct magnetic phases within a material.

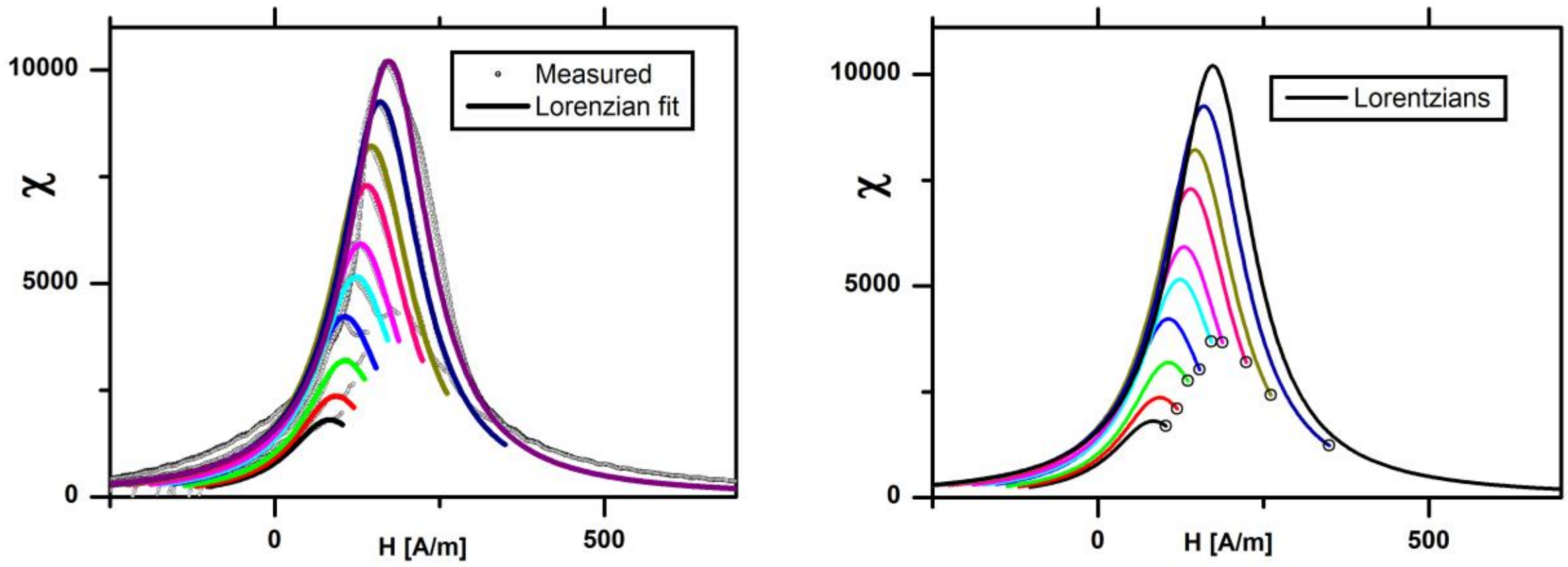

**FIG.6.** Minor differential susceptibility curves and corresponding Lorentzians for a low-carbon steel.

Table 5. Lorentzian parameters for the minor differential susceptibility curves. $H_{MAX}$ is the minor loop amplitude.

| $H_{MAX}$ (A/m) | $(2/\pi)\ a\ M_{MAX}$ | $1/a$ (A/m) | $H_C$ (A/m) |
|---|---|---|---|
| 103 | 1808 | 70 | 83 |
| 119 | 2364 | 70 | 93 |
| 136 | 3192 | 70 | 107 |
| 153 | 4222 | 70 | 107 |
| 170 | 5156 | 70 | 124 |
| 188 | 5924 | 70 | 130 |
| 223 | 7292 | 70 | 140 |
| 260 | 8217 | 70 | 147 |
| 350 | 9252 | 70 | 160 |
| 880 | 10202 | 70 | 173 |

## III. Discussion

We see that the Lorentzian function describes DS of polycrystalline steels with very high accuracy for non-oriented materials and for grain-oriented ones. Also it gives a very good agreement with experimental curves of quenched and tempered steels [31]. In the polycrystalline material due to a grain misorientation from the field direction there is the demagnetizing field inside the grain proportional to the grain magnetization. Let us consider a collection of volumes/grains with the magnetization curve *M(H)* given by the demagnetizing factor and all demagnetizing factors being equally distributed. We assume that in each grain the internal magnetization curve (no demagnetizing fields) is rectangular – the susceptibility approaches the delta-function. Let us consider eight grains having the same high constant differential susceptibility $\chi$ =128 in the field range ±1. Let us take the first grain with no demagnetizing field. The susceptibility in the second grain due to the demagnetizing field is 64 in the field range ±2, in the third – 32 in the field range ±4 and so on. If we look at the sum of all susceptibilities, we see that the total susceptibility is close to the Lorentzian shape (see Fig.7). Actually, this series follows the $(1/2)^{0.5x}$ function, which is very close to the Lorentzian (see Fig.8).

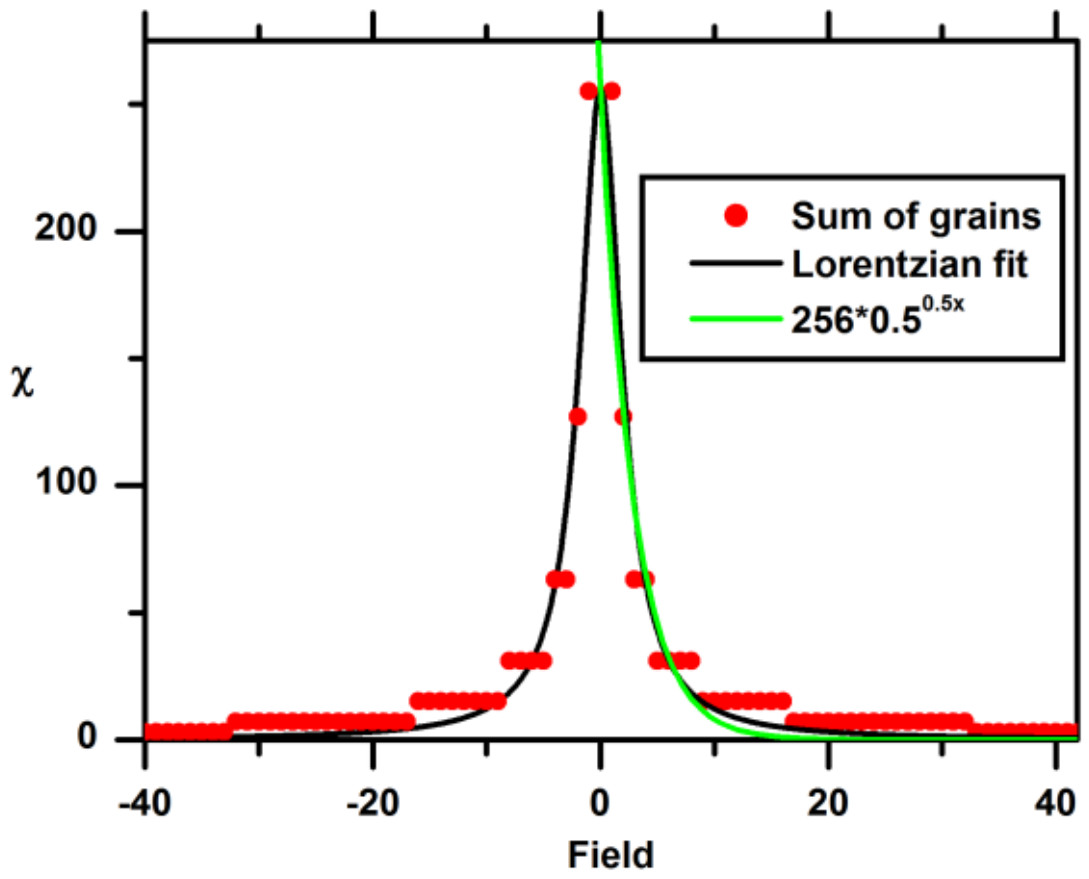

**FIG.7.** The differential susceptibility by a collection of eight grains with different demagnetizing factors.

The considerations above are just illustrative, but they give a clue how the one-peak DS curve is obtained for a polycrystalline material. To understand why *M(H)* curves of such materials follow the arctangent shape with a very high accuracy, 2D and 3D statistical calculations of collections of grains are necessary for different grain angle distributions. The problem of *M(H)* calculation of a collection of differently oriented magnetostatically interacting grains is extremely complex. Even for much simpler systems with interacting arrays of equally spaced magnetic moments (spins) there are no analytical solutions. The only case when one can come to an analytical solution is the collection of noninteracting particles (grains) with a certain distribution of susceptibilities. It is amazing and completely unexpected that the *M(H)* curve of such a very complex system is a simple analytical function.

The simplest polycrystalline material, for which the Lorentzian describes the DS with a high accuracy (see Fig.3) is a Goss-textured Fe-3%Si thin sheet. The average easy axis is along the field direction with small grains misorientation angles. The magnetic domains are oriented in each grain along the easy [001] axis closest to the applied magnetic field direction. Due to the grain misorientation demagnetizing fields appear at the grain boundaries and at the sheet surface. Supplementary domains minimize these demagnetizing fields. Removal of the supplementary domains by tensile stress (see Fig.8) increases demagnetizing fields that substantially affects the *M(H)* loops shape [33]. So even in this simplest case one has to include the supplementary (closure) domains and corresponding demagnetizing fields to create a realistic micromagnetic statistical model of a magnetization curve. The magnetic domains around a misoriented grain and the finite elements method calculations of the field across such a grain are shown in Fig.9.

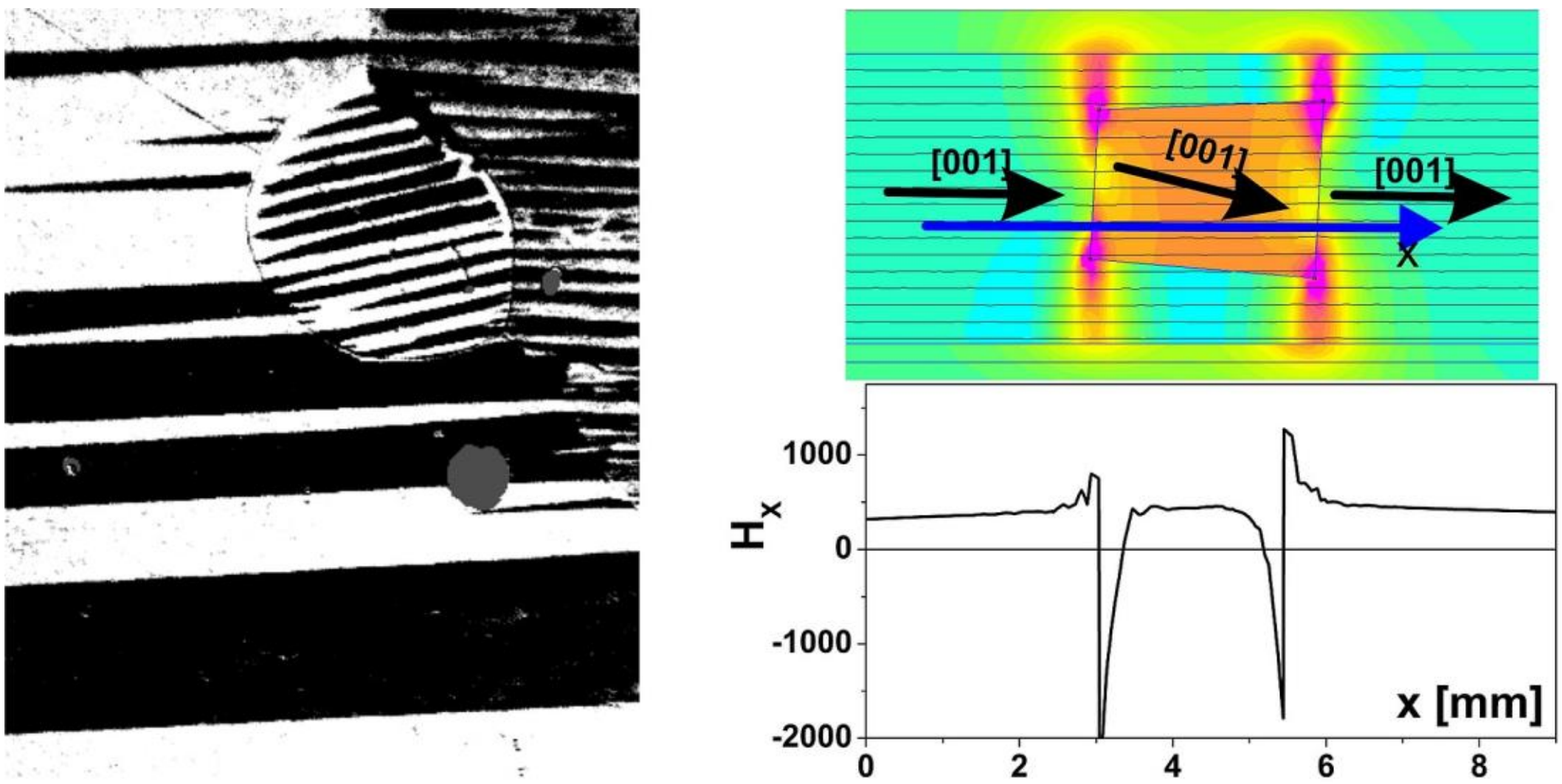

**FIG.8.** Magnetic domains around a misoriented grain under 50 MPa tensile stress and finite elements method (FEMM 2D) calculation of the internal field distribution across such a grain. The applied field is horizontal, close to the easy magnetization direction of grains.

It is intriguing that the DS peaks of elastically and plastically deformed materials measured along the magnetically harder direction are also a sum of Lorentzians (see Fig.2 and Fig.3). We have shown that two- and three-phase hysteresis loops could be fitted by two and three Lorentzians correspondingly with very good agreement with experimental data. The arctangent function for the magnetization curve has three parameters - the saturation magnetization, the coercivity and the magnetization slope (maximum differential susceptibility) at the coercivity. It is a simplest function with a minimum necessary number of physical parameters that surprisingly give a very good agreement with experimental curves of polycrystalline materials. Another surprising fact is that the susceptibility is usually decomposed into the reversible and irreversible part [3-7], while the Lorentzian describes the total susceptibility ignoring separation into individual components.

One has to keep in mind that the model should be applied to $M(H)$ of a material measured accurately. It is not an easy task for a yoke-sample setup if air gaps are present that produce demagnetizing fields. The toroidal or ring-shaped samples are closed magnetic circuit, but even in this "ideal" case there can be large measurement errors in $M$ and $H$ if the ratio of the sample thickness to the diameter is lower than 1 to 10 [35].

## IV. Conclusions

We see that the Lorentzian, which is commonly used in spectroscopy to fit the data to extract individual components could be used in the same way to extract individual magnetic components from the differential susceptibility curves. All the necessary algorithms and procedures could be taken from the optical and X-ray diffraction spectroscopy. The magnetic spectroscopy using Lorentzians could especially helpful for analysis of individual phases in multi-phase materials. Often there is a second magnetically harder phase that is hardly seen on the total curve, which by the magnetostatic interaction with the main softer phase makes the DS curve to shift to higher fields with the field amplitude increase also making the curve more and more asymmetric [26]. So, from the deviation of DS curve from the Lorentzian shape one can detect the presence of the second phase. Sometimes, the second phase is close magnetically to the first one and the DS curves of two phases are very close and can't be easy distinguished by the naked eye like the deformed layer of magnetically hard bearing steels [34]. In this case the analysis based on the search of a sum of Lorentzian peaks will be helpful. In the case of DS of a materials under stress the description of magnetic properties change is reduced to just a few parameters of one or two Lorentzians.

## Declaration of competing interest

The authors have no conflicts to disclose.

## Data availability

The data that support the findings of this study are available from the corresponding authors upon request. The data are also available in Zenodo at https://doi.org/10.5281/zenodo.15209641 [36]. Datasheets of hysteresis loops under stress can be found in [37].

## Acknowledgments

The authors acknowledge the assistance provided by the Ferroic Multifunctionalities project, supported by the Ministry of Education, Youth, and Sports of the Czech Republic. Project No. CZ.02.01.01/00/22_008/0004591, co-funded by the European Union.